\documentclass{article}
\usepackage[utf8]{inputenc}
\usepackage{pdfpages}  
\usepackage[T1]{fontenc}       
\usepackage[T1]{fontenc}
\usepackage{pgfplots}
\usepackage{amsmath}
\usepackage{verbatim}

\DeclareMathOperator{\er}{\vec{e}_r}
\DeclareMathOperator{\et}{\vec{e}_\alpha}
\usepackage{tikz}
\usepackage{graphicx}
\title{Liquid walls and interfaces in arbitrary directions stabilized by vibrations \\ \* \\
Supplementary informations
}
\author{Benjamin Apffel, Samuel Hidalgo-Caballero, \\ Antonin Eddi, Emmanuel Fort \\}
\date{Corresponding author : emmanuel.fort@espci.fr}
\begin{document}



# Liquid walls and interfaces in arbitrary directions stabilized by vibrations

Benjamin Apffel[1], Samuel Hidalgo-Caballero[1], Antonin Eddi[2], Emmanuel Fort[1,*]

[1]ESPCI Paris, PSL University, CNRS, Institut Langevin, 1 rue Jussieu, F-75005 Paris, France.

[2] PMMH, CNRS, ESPCI Paris, Université PSL, Sorbonne Université, Université de Paris, F-75005, Paris, France

*Correspondence to: emmanuel.fort@espci.fr

**Abstract**

Gravity shapes liquids and play a crucial role in their internal balance. Creating new equilibrium configurations irrespective of the presence of a gravitational field is challenging with applications on earth as well as in zero-gravity environments. Vibrations are known to alter the shape of liquid interfaces and to also to change internal dynamics and stability in depth. Here, we show that vibrations can also create an "artificial gravity" in any direction. We demonstrate that a liquid can maintain an inclined interface when shaken in an arbitrary direction. A necessary condition for the equilibrium to occur is the existence of a velocity gradient determined by dynamical boundary conditions. However, no-slip boundary condition and incompressibility can perturb the required velocity profile leading to a destabilization of the equilibrium. We show that liquid layers provide a solution and liquid walls of several centimeters in height can thus be stabilized. We show that the buoyancy equilibrium is not affected by the forcing.

Vibrations can have a dramatic effect on a fluid equilibrium inducing a slow global reorientation of the fluid toward a new quasi-static equilibrium. Faraday was the first to observe this phenomenon called vibro-equilibrium in 1831 with the flattening of a droplet submitted to vertical vibrations[1]. Since then, the effects of vibrations on the shape and stability of liquid interfaces have been studied in various configurations[2,3]. In the case of vertical oscillations, it can lead either to stabilization as for upside-down interfaces with the suppression of the Rayleigh-Taylor instability[2,4,5] or to instability with the parametric excitation of waves (Faraday instability)[1,6,7]. In the horizontal direction, the interface shape differ from the unforced equilibrium with the excitation of frozen-like waves[2,8,9]. Vibroequilibria and interface stabilization have been analyzed starting from its solid analog with the Kapitza's pendulum model[10]. This approach is based on the decoupling of the dynamics into a fast motion at the vibration frequency and a slow time-averaged motion. Various variational approaches with functionals including terms associated with potential energy as well as vibrational energy also have been developed to obtain a more precise description of the shape of the interface[11–14].

Beyond shaping the liquid interface, vibrations have also a strong effect on the dynamics and equilibria taking place inside the fluids or heterogeneous systems. It has been shown that vibrations can promote mixing and favor chemical reactions[15,16], induce Bjerknes forces[17] on bubbles and immersed bodies[18–23] or stabilize upside-down buoyancy equilibria[24].

In most typical situations, the effect of vibrations is negligible compared to the more familiar effects of gravity. Many studies have thus been performed in microgravity[25]. Beyond interface modifications, vibrations induce changes in the fluid dynamics creating an "artificial gravity" in space[26–29].

Here, we investigate the ability for strong vibrations to create a controlled "artificial gravity" in an arbitrary direction in the presence of gravitation. We show that slanted liquid interfaces can be stabilized in all directions. The vertical pressure gradients due to gravity are balanced by a kinematic pressure term created by a velocity gradient at the interface. Tilted liquid baths are limited by volume conservation while liquid walls surrounded by compressible air layers can be stabilized by lifting this non-compressibility constraint. Buoyancy equilibria are recovered on each interface.

The experimental setup is composed of a plexiglass container mounted on an electromagnetic shaker. The whole system can be rotated freely to change the angle of the excitation $\theta$ to the

vertical axis. All the experiments were performed with silicon oil with dynamic viscosity $\eta = 1000$ cSt and density $\rho = 920$ kg/m$^3$.

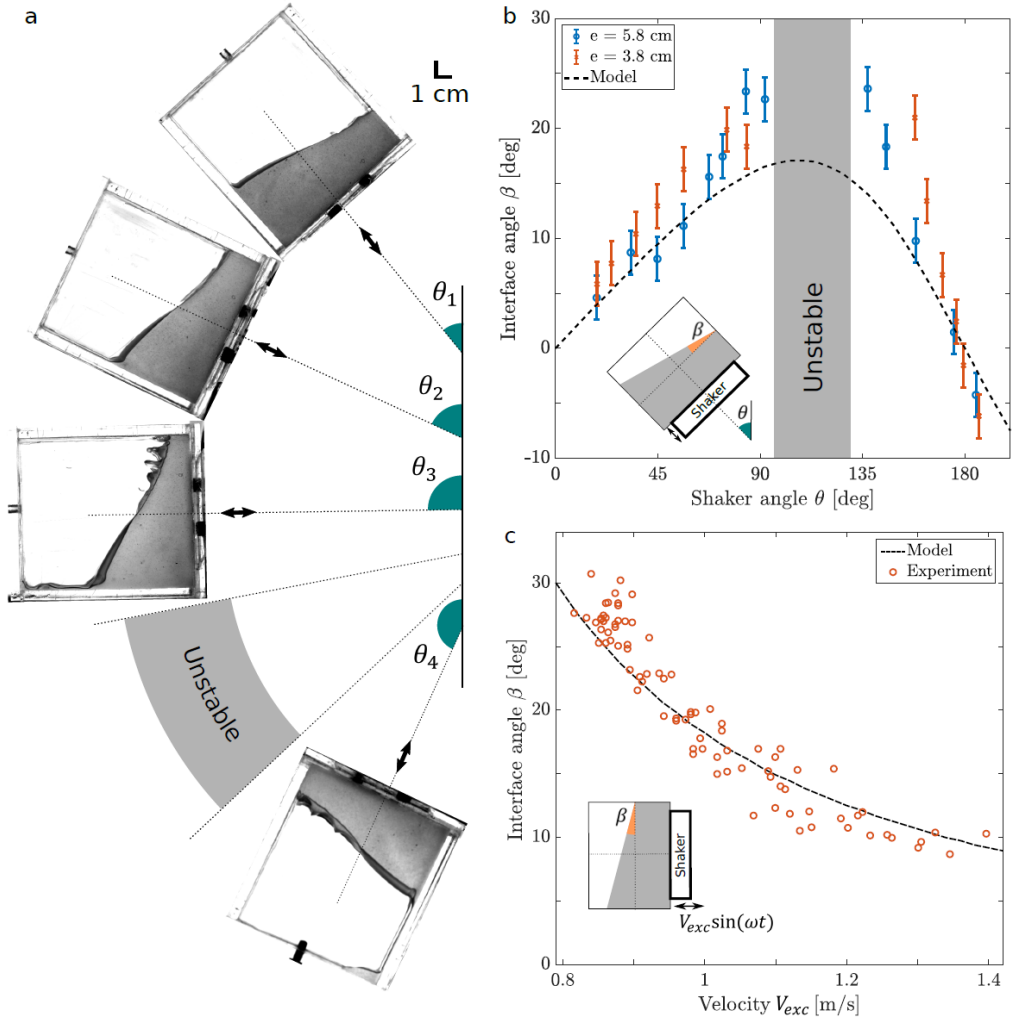

**Figure 1. Liquid vibro-equilibria in arbitrary forcing directions. a**, Snapshots of the liquid container of width $L = 9$ cm shaken in at various angles $\theta$ (Supplementary Video 1). The liquid is silicon oil with dynamic viscosity $\eta = 1000$ cSt and its height is $e = 3.4$ cm. The forcing is at frequency $\omega/2\pi = 100$ Hz and velocity amplitude $V_{exc} = 1.4$ m.s$^{-1}$. **b**, Angle $\beta$ between the liquid interface to the forcing direction (inset) as a function of the excitation angle $\theta$ for two liquid depths $e = 3.8$ cm (red crosses) and $e = 5.8$ cm (blue circles) with the same forcing. The error bars were evaluated experimentally at $\pm 2°$. The Kapitza's pendulum model (dashed line) is taken with $\omega_0 \sim \sqrt{\pi g/L} = 18$ rad/s (Supplementary Information). **c**, Angle $\beta$ as a function of the forcing velocity amplitude $V_{exc}$ in a container of width $L = 5$ cm. Experimental data (circles) and Kaptiza's pendulum model

(dashed lines) in the same experimental conditions as **a** for an horizontal shaking direction $\theta = 90°$.

We first study the influence of the excitation angle on the liquid vibroequilibrium. Figure 1a shows snapshots of the container of width $L = 9$ cm filled with a liquid depth $e = 3.8$ cm for several excitation angles $\theta$. The vibration frequency is chosen at $\frac{\omega}{2\pi} = 100$ Hz and the container velocity amplitude $V_{\text{exc}} = 1.4$ m/s. The liquid interface is approximately flat and can be characterized by its average angle $\beta$ between the normal to the interface and the direction of vibration (Fig. 1b). This angle is slowly increasing from $\beta = 0$ in the standard upright position ($\theta = 0°$) to approximately $\beta = 20°$ at the angle $\theta = 100°$ for which the liquid interface is approximately vertical. Upon further increasing the excitation angle, the interface destabilizes, leading to the fall of the liquid. However, since it takes a finite time for the liquid to destabilize, it is possible to explore higher angles $\theta$ by a fast rotation the container. An equilibrium is recovered above $\theta \sim 130°$ up to the upside-down position ($\theta = 180°$). The angle $\beta$ decreases in this interval from 20° to 0°. The variation of $\beta$ is independent of the liquid depth for sufficient large depths (Fig. 1b).

When an equilibrium is observed, the vibration induces a reorientation of the liquid toward a new quasi-static equilibrium. The periodic oscillations at the liquid surface set by the forcing frequency on the fast timescale drive a slower reorientation of the fluid toward a new quasi-steady equilibrium distinct from the hydrostatic one oriented toward the direction of the excitation[30]. This resemble the Kapitza's pendulum for which a pendulum is vibrated vertically and stabilized in the upside-down position with, in this case, a forcing in an arbitrary direction $\theta$. The dynamic of the pendulum can also be separated in two time scales[10]. The angle $\beta$ between the pendulum of angular frequency $\omega_0$ and the forcing direction $\theta$ when shaken with a velocity $v_{\text{exc}}(t) = V_{\text{exc}} \cos(\omega t)$ in the direction $\theta$ can be written as the sum of a fast component $\beta_f$ at the forcing frequency and a slow averaged one $\beta_s$. For a strong forcing, $\beta_s$ takes the simple following form (see general case in Supplementary Information)

$$\ddot{\beta}_s + \omega_1^2 \beta_s = \omega_0^2 \sin \theta \qquad \text{with} \qquad \omega_1^2 = \omega_0^2 \left[ \cos \theta + \frac{V_{\text{exc}}^2 \omega_0^2}{2g^2} \right] \tag{1}$$

and $g$ being the acceleration of gravity. The average equilibrium angle is thus given by $\beta_{\text{eq}} = \frac{\omega_0^2}{\omega_1^2} \sin \theta$ which defines two diametrically opposed equilibrium angles. The observed

experimental solution results from a slow variation of $\theta$. The analog of the pendulum's angular frequency $\omega_0$ for the liquid must be determined. The scaling should be $\omega_0 \sim \sqrt{g/L_i}$ with $L_i$ being the characteristic length of the fluid interface. For small $\beta_{eq}$, $L_i \sim L$ taking the expression given by the wave dispersion relation $\omega_0 \sim \sqrt{\pi g/L}$. This value gives a very good agreement with the measured angles $\beta_{eq}$ as a function of the excitation direction $\theta$ without any fitting parameter (Fig. 1b). In addition, the variations of $\beta_{eq}$ with the forcing velocity $V_{exc}$ for a horizontal excitation ($\theta = 90°$) also agrees very well with the model (Fig. 1c). Upon increasing the excitation, the fluid surface reorient itself in the orthogonal direction.

In the case of the pendulum, the equilibrium is stable if $\omega_1^2 > 0$ and only inverted pendulums ($\cos\theta < 0$) can be unstable. The instability increases with the increasing downward direction but all directions can be stabilized upon sufficient forcing above a critical speed $V_K^* \sim \sqrt{2gL_i/\pi}$. In liquid analog, the instability also occurs for inverted interfaces but the stabilization is recovered in the lower inverted quadrant ($\theta > 130°$)[24,31].

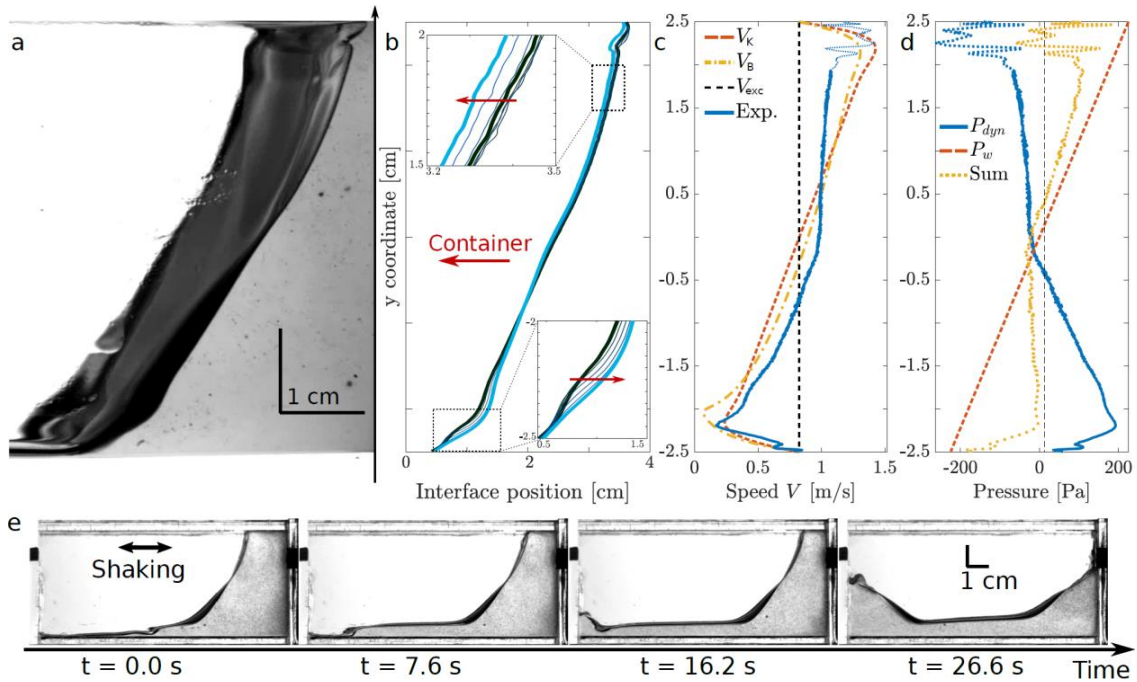

**Figure 2. Dynamic and stability of the slanted interface. a**, Typical snapshot of the liquid shaken in the horizontal direction $\theta = 90°$ **b**, Profiles of the liquid interface measured from images like in **a** at various times during a forcing period (Supplementary Video 2). The interface is oscillating in phase with the container for the upper part (see close-up) and $\pi$-shifted for the lower part (see close-up) at the forcing frequency in the container frame. The

two thick lines are associated to the extremal positions of the container. **c**, Experimental profile of the velocity amplitude $V_{\text{exp}}$ as a function of the height $y$ in the lab frame deduced from **b** (solid blue curve) together with the one obtained from Kapitza's model $V_K$ (dashed red line) and that from the dynamical model $V_B$ (dashed yellow line). The velocity amplitude of the container is $V_{\text{exc}} = 0.8$ m/s (vertical dotted Z line). The upper part of the curve is noisy due to the presence of Faraday waves. The profiles obtained from the models are linked with the non-slip conditions using the Stokes equation (Supplementary Information) **d**, Pressure contributions along the interface as a function of the height $y$: hydrostatic pressure $P_w = -\rho g y$, experimental dynamic pressure $P_{\text{dyn}} = \frac{1}{2}\rho\left(V_{\text{exp}}/\sqrt{2}\right)^2$ and the sum $P_w + P_{\text{dyn}}$ which must be approximately constant at equilibrium. **e**, Snapshots of the drainage of the liquid in a container of width $L = 5$ cm shaken in the horizontal direction (Supplementary Video 3). The time interval between 2 snapshots lasts several seconds. The liquid reaches a final symmetric vibro-equilibrium state shown in the last snapshot.

In contrast to rigid pendulums, the forcing must also ensure the equilibrium of the interface in the liquid case. The slanted liquid interface must be stabilized. Figure 2a shows a typical snapshot of the liquid vibrated horizontally. It is possible to extract the fast dynamics of the interface profile in the oscillating frame (Fig. 2b). The interface undergoes a global oscillation at the excitation frequency locked in phase with the forcing. The lower part of the interface being $\pi$-shifted while the upper part is in phase with a fixed point just below the middle of the interface's height. By observing the motion of small particles, the velocity near the interface is found to be mostly normal to the surface (Supplementary Videos 2 and 3). In the lab frame, the amplitude of the velocity $V_{\text{exp}}(y)$ at the forcing frequency can thus be deduced by simply looking at the motion of the interface. This amplitude is increasing with the height $y$ (Fig. 2c). At the boundaries, a no-slip condition imposes an inversion of the velocity gradient over a boundary layer of a few millimeters. The presence of Faraday waves in the upper part of the interface is a signature of the higher velocity amplitude as the instability threshold is crossed (Fig. 1a). The same type of velocity profiles are observed for all the liquid equilibria with slanted interfaces whatever the excitation direction.

The stability of the slanted interface can be understood by the inhomogeneous velocity amplitude along the interface which provides a kinetic potential to counteract the hydrostatic equilibrium. This dynamical term appears by writing that the pressure should be constant at the interface. The correlation between the fast oscillating pressure in the fluid and the interface

motion generates an extra stabilizing term that depends on the interface velocity[8]. In the large forcing regime, the equilibrium satisfies (Supplementary Information)

$$\frac{d}{dy}\left(\frac{1}{2}<v_B(y,t)^2>-gy\right)=0 \quad (2)$$

with $<.>$ being the time averaging operation over the forcing period of the instantaneous velocity $v_B(y,t)$ at time $t$ and height $y$. The velocity amplitude at the forcing frequency thus satisfies $V_B(y) = V_{exc}\sqrt{1+4g(y-y_0)/V_{exc}^2}$ with $y_0$ being a fixed point in the container's frame. The position of the fixed point is given by the constraint imposed by volume conservation. In addition, because of the non-slip condition at the interface, the velocity profile must be modified near the walls. Using the Stokes equation in the co-moving frame for the boundary layer (Supplementary Information), we obtain the velocity profile plotted in Fig. 2c which is in good agreement with the experimental one. This analysis yields a critical excitation velocity for a given interface height required by the existence of a minimum velocity along the entire profile.

It is interesting to compare this result with the one obtained from Kapitza's pendulum. In this latter case, the pendulum is oscillating at the forcing frequency around an equilibrium position $\beta_{eq}$. For an intense forcing ($\cos\theta \ll \frac{V_{exc}^2\omega_0^2}{2g^2}$), the instantaneous velocity $v_K(y,t)$ along a plane orthogonal to the pendulum coincident with that of the liquid interface satisfies $v_K(y,t) = V_K\cos\omega t$ with $V_K = V_{exc} + 2g(y-y_0)/V_{exc}$ and $y_0 = L/2$ (Supplementary Information). Figure 2c shows the linear profile associated to the average value of the velocity with Stokes-type linking for the non-slip condition to the boundaries in the case of a horizontal shaking. This profile is in good agreement with the experimental data. It corresponds to the first-order approximation of the non-linear velocity profile deduced from the boundary condition at the interface. The velocity gradient results from the phase-locked oscillation of the pendulum with its forcing motion. Figure 2d shows the various contributions to the pressure along the interface as a function of the height $y$. The sum of the hydrostatic pressure $P_w = -\rho g y$ and the dynamic pressure $P_{dyn} = \frac{1}{2}\rho(V_{exp}/\sqrt{2})^2$ are nearly constant along the interface as expected at equilibrium.

At the bottom of the interface the model imposes $V_B < V_{exc}$ from volume conservation while the no-slip boundary condition imposes $V_B = V_{exc}$ at the wall. As a result, the velocity gradient along the interface changes sign (Fig. 2c), breaking the equilibrium condition. The liquid is drained slowly at the lower interface in the shape of a liquid finger of a few millimeters in

height (Fig. 2e and Supplementary Video 3). Hence, interface effects produce an unstable range of excitation angles around $100° < \theta < 130°$. Surface tension provides a stabilizing effect on the liquid drainage over the capillary length size at the interface. The drainage seems to be triggered when the typical width of the inversed velocity gradient layer exceeds the capillary length. This agrees with the observed instability angles $\theta$ for which the height of the liquid interface is the largest. Hence, in contrast to solid pendulums, the stability of vibrated liquids is hindered by non-slip boundary effects in forcing directions associated to the largest liquid heights.

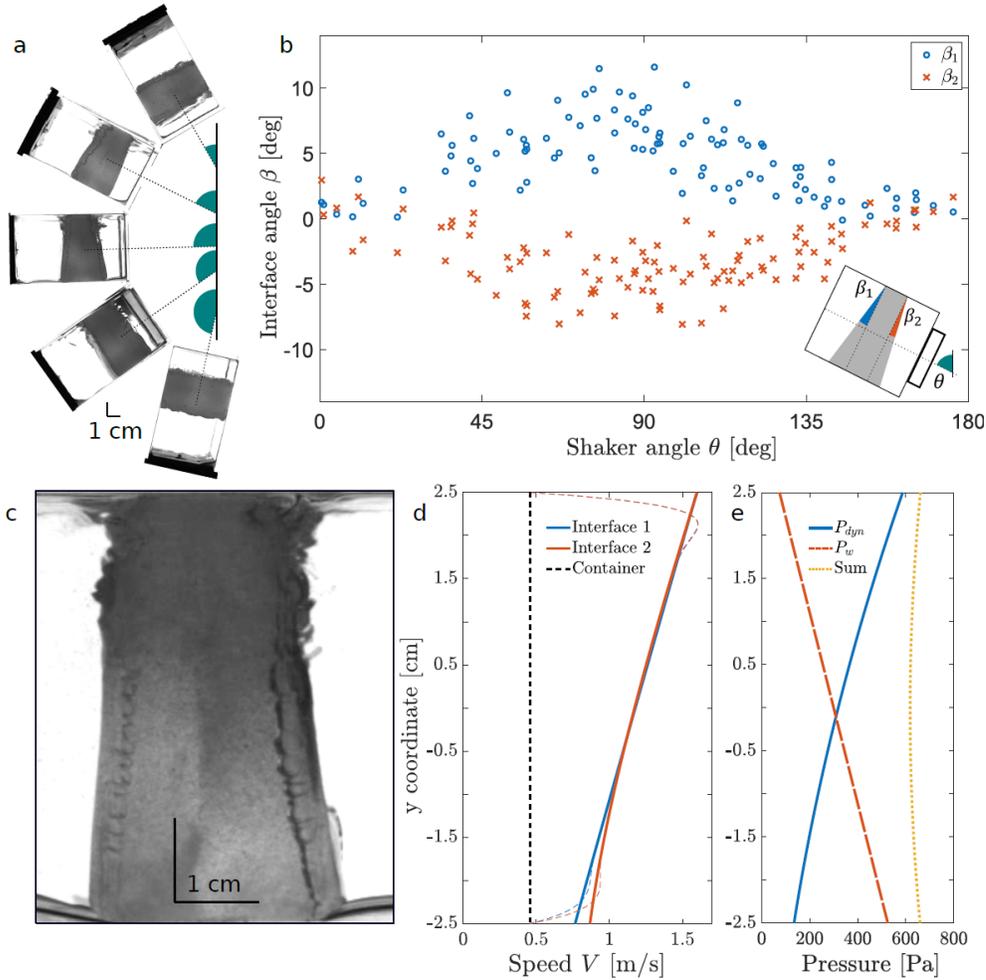

**Figure 3. Vibro-equilibrium of liquid walls. a**, Snapshots of a liquid layer of thickness $e =$ and width $L =$ for various excitation angles $\theta$ (Supplementary Video 4). The liquid layer is stable in all the directions. The excitation is set at a frequency of 100 Hz and a velocity amplitude of the container is $V_{\text{exc}} = 0.45$ m/s while the velocity of the layer is 1 m/s. **b**, Angles $\beta_1$ and $\beta_2$ between each interface of the liquid layer and the forcing direction (see

inset) as a function of the latter $\theta$. **c**, Typical snapshot of a vertical liquid wall stabilized by horizontal forcing $\theta = 90°$. **d**, Profile of the velocity amplitude $V_{exp}$ for the two interfaces (solid lines) as a function of the height $y$ in the lab frame compared with the container excitation $V_{exc}$. These curves are calculated from the oscillations of the interface profiles observed in snapshots acquired at a different time of the forcing period (Supplementary Video 5). Because of the presence of Faraday waves and of the inward meniscus orientation, the interface profiles were approximated by linear fits. Taking into account the non-slip condition modifies the velocity amplitude profile at the boundary (colored dashed lines). **e**, Pressure contributions along the interface as a function of the height $y$: hydrostatic pressure $P_w = \rho g y$, experimental dynamic pressure $P_{dyn} = \frac{1}{2}\rho(V_{exp}/\sqrt{2})^2$ and the sum $P_w + P_{dyn}$ which must be constant at equilibrium.

The destabilizing change of sign of the velocity gradient at the boundary is imposed by volume conservation which is a consequence of the non-compressibility of the liquid. A liquid layer with compressible air on both sides can alleviate this constraint. Such liquid layers can be obtained by injecting air with a needle at the bottom of a vibrating container filled with liquid. Since bubbles can sink into a shaken liquid, an air cushion can be created to support the liquid[18–21,24]. This liquid layer appears stable for all the excitation angles when the shaker is rotated (Fig. 3a and Supplementary Video 4). The measured tilt angles of the two interfaces $\beta_1$ and $\beta_2$ as a function of the excitation angle $\theta$ (Fig. 3b) show the same symmetric smooth shape increasing in absolute value from zero at $\theta = 0°$ to a maximum value at $\theta = 90°$ and decreasing back to zero at $\theta = 180°$. Note that the stability of the liquid layer is increased experimentally by closing the two sides of the container with just one pinhole for the injection needle to pass through. The two compressible air layers act as springs to excite the liquid layer which has a resulting average velocity $V_{liq}$ that can be significantly larger than that of the container itself[24]. For the dataset shown in Fig. 3, we measured $V_{exc} = 0.45$ m/s and $V_{liq} = 1.0$ m/s.

Due to the relaxation of the volume conservation in the air layers, the liquid layer can now undergo some deformation during an excitation period which make the analogy with a solid pendulum less pertinent. The variations of $\beta$ with the excitation direction $\theta$ are similar to that of the filled container for the stable excitation directions (Fig. 1). However, the amplitudes of the tilt angles appear much smaller in the former case, even if $V_{liq} > V_{exc}$ is taken into account

for the forcing. The velocity profile should still satisfies the dynamical equation (2) to fulfill the stability condition of the liquid interface. The interface motion can be obtained from side views of a vertical liquid layer of 5 cm in height shaken horizontally (Fig. 3c and Supplementary Video 5). To obtain the velocity profile $V_{exp}(y)$, each interface profile is approximated by a linear fit due to the presence of the Faraday waves in the upper part and to the inward meniscus orientation which prevent a more detailed analysis (see solid lines in Fig. 3d). The non-slip condition at each boundary is again taken into account with Stokes equations (colored dashed lines). The velocity amplitude $V_{exp}(y)$ of liquid walls is found always higher than the velocity amplitude of the container $V_{exc}$ (dashed lines). In addition, the velocity profile increases continuously up to the upper boundary layer insuring the stability of the layer. Note that the two interfaces have nearly equal velocity profiles as expected from the incompressibility of the liquid layer.

Pressure contributions along the interface as a function of the height $y$ (Fig 3e) show that the sum of the hydrostatic pressure $P_w = \rho g y$ and the experimental dynamic pressure $P_{dyn} = \frac{1}{2}\rho\big(V_{exp}(y)/\sqrt{2}\big)^2$ is approximately constant as needed at equilibrium. Unlike the filled container case, the pressure stability is now insured by the absence of inversion of the velocity gradient profile $V_{exp}(y)$ in the lowest boundary. The presence of a compressible air layer in liquid walls is necessary to achieve this pressure stability.

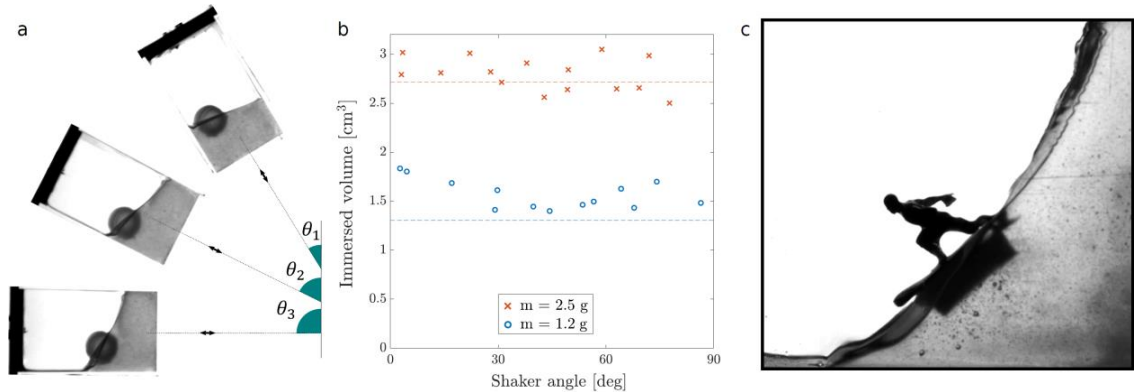

**Figure 4 Buoyancy on slanted liquid interfaces. a**, Snapshots of plastic sphere floating on a vibrated liquid at various excitation angles $\theta$ (Supplementary Video 6). The fluid is silicon oil shaken at 100 Hz and with a velocity amplitude $V_{exc} =$. **b**, Experimentally measured immersed volume as a function of the forcing angle $\theta$ for 1.5 cm-diameter spheres with mass 2.5 g (red crosses) and 1.2 g (blue circles). Dashed lines represent the expected immersed

volume from classical buoyancy with no fitting parameter. **c**, 3D printed surfer of approximately 2 cm in height at equilibrium on a stable wave for a liquid shaken horizontally (Supplementary Video 7).

It has been previously shown that vertical shaking could stabilize unstable equilibrium positions of objects floating upside-down[24]. How does shaking affects the buoyancy equilibrium in an arbitrary direction? For a filled container, the floating position is measured as a function of the excitation angle $\theta$ for 1.5 cm-diameter spheres of various weights (see snapshots Fig. 4a and Supplementary Video 6). The immersed part of the sphere remains unperturbed for a given object density as the shaking is rotated (Fig. 4b) showing that the buoyancy equilibrium is not changed by the forcing and remains normal to the iso-pressure interface. Note that secondary flows in the liquid have a slight influence on the transverse position of the floaters which tend to drift slowly into preferred positions. The floating objects behave as if a gravitational force of constant amplitude was exerted perpendicular to the liquid surface whatever the excitation direction. The shaking thus creating a new vibro-equilibrium in a sort of "artificial gravity" induced by the dynamic energy provided by the shaking. Hence, in this environment, surfer toys can surf on a customized stable wave (Supplementary Video 7).

Vibrations can create new dynamical equilibria in an arbitrary direction in which the effect of gravity can be neglected. This offer new opportunities to control more complex interactions and for new unexplored configurations involving fluid mixture dynamics, in zero gravity environments or more complex multiphase systems.

## Acknowledgments

## References

1. Faraday, M. On a Peculiar Class of Acoustical Figures; and on Certain Forms Assumed by Groups of Particles upon Vibrating Elastic Surfaces. *Philosophical Transactions of the Royal Society of London* **121**, 299–340 (1831).

2. Wolf, G. H. The dynamic stabilization of the Rayleigh-Taylor instability and the corresponding dynamic equilibrium. *Z. Physik* **227**, 291–300 (1969).

3. Wolf, G. H. Dynamic Stabilization of the Interchange Instability of a Liquid-Gas Interface. *Phys. Rev. Lett.* **24**, 444–446 (1970).

4. Lapuerta, V., Mancebo, F. J. & Vega, J. M. Control of Rayleigh-Taylor instability by vertical vibration in large aspect ratio containers. *Phys. Rev. E* **64**, 016318 (2001).

5. Sterman-Cohen, E., Bestehorn, M. & Oron, A. Rayleigh-Taylor instability in thin liquid films subjected to harmonic vibration. *Physics of Fluids* **29**, 052105 (2017).

6. Kumar, S. Parametric instability of a liquid sheet. *Proceedings of the Royal Society A: Mathematical, Physical and Engineering Sciences* **457**, 1315–1326 (2001).

7. Douady, S. Experimental study of the Faraday instability. *J. Fluid Mech.* **221**, 383–409 (1990).

8. Lyubimov, D. V. & Cherepanov, A. A. Development of a steady relief at the interface of fluids in a vibrational field. *Fluid Dyn* **21**, 849–854 (1986).

9. Talib, E., Jalikop, S. V. & Juel, A. The influence of viscosity on the frozen wave instability: theory and experiment. *Journal of Fluid Mechanics* **584**, 45–68 (2007).

10. Kapitza, P. L. Dynamic stability of a pendulum when its point of suspension vibrates. *Soviet. Phys. JETP* **21**, 588–597 (1951).

11. Lukovskii, I. A. & Timokha, A. N. Variational formulations of nonlinear boundary-value problems with a free boundary in the theory of interaction of surface waves with acoustic fields. *Ukr Math J* **45**, 1849–1860 (1993).

12. Blekhman, I. I. *Vibrational Mechanics: Nonlinear Dynamic Effects, General Approach, Applications*. (WORLD SCIENTIFIC, 2000). doi:10.1142/4116.

13. Beyer, K., Gawriljuk, I., Günther, M., Lukovsky, I. & Timokha, A. Compressible Potential Flows with Free Boundaries. Part I: Vibrocapillary Equilibria. *ZAMM - Journal*

*of Applied Mathematics and Mechanics / Zeitschrift für Angewandte Mathematik und Mechanik* **81**, 261–271 (2001).

14. Gavrilyuk, I., Lukovsky, I. & Timokha, A. Two-dimensional variational vibroequilibria and Faraday?s drops. *Z. angew. Math. Phys.* **55**, 1015–1033 (2004).

15. Ito, Y. & Komori, S. A vibration technique for promoting liquid mixing and reaction in a microchannel. *AIChE Journal* **52**, 3011–3017 (2006).

16. Elbing, B. R., Still, A. L. & Ghajar, A. J. Review of Bubble Column Reactors with Vibration. *Ind. Eng. Chem. Res.* **55**, 385–403 (2016).

17. Bjerknes, V. F. K. *Fields of Force: Supplementary Lectures, Applications to Meteorology*. (Columbia University Press/The Macmillan Company, 1906).

18. Baird, M. H. I. Resonant bubbles in a vertically vibrating liquid column. *The Canadian Journal of Chemical Engineering* **41**, 52–55 (1963).

19. Sorokin, V. S., Blekhman, I. I. & Vasilkov, V. B. Motion of a gas bubble in fluid under vibration. *Nonlinear Dyn* **67**, 147–158 (2012).

20. Blekhman, I. I., Blekhman, L. I., Vaisberg, L. A., Vasil'kov, V. B. & Yakimova, K. S. "Anomalous" phenomena in fluid under the action of vibration. *Dokl. Phys.* **53**, 520–524 (2008).

21. Blekhman, I. I., Blekhman, L. I., Sorokin, V. S., Vasilkov, V. B. & Yakimova, K. S. Surface and volumetric effects in a fluid subjected to high-frequency vibration. *Proceedings of the Institution of Mechanical Engineers, Part C: Journal of Mechanical Engineering Science* **226**, 2028–2043 (2012).

22. Zen'kovskaja, S. M. & Novosjadlyj, V. A. Influence of vertical oscillations on bilaminar system with a non-rigid interface , vol. 48, no. 9, pp 1710-1720. *Zhurnal vychislitel'noj matematiki i matematicheskoj fiziki* **48**, 1710–1720 (2008).

23. Chelomei, V. N. Mechanical paradoxes caused by vibrations. *Soviet Physics Doklady* **28**, 387 (1983).

24. Apffel, B., Novkoski, F., Eddi, A. & Fort, E. Floating under a levitating liquid. *Nature* **585**, 48–52 (2020).

25. Salgado Sánchez, P., Fernández, J., Tinao, I. & Porter, J. Vibroequilibria in microgravity: Comparison of experiments and theory. *Phys. Rev. E* **100**, 063103 (2019).

26. Beysens, D. Vibrations in space as an artificial gravity? *Europhysics News* **37**, 22–25 (2006).

27. Lyubimova, T., Ivantsov, A., Garrabos, Y., Lecoutre, C. & Beysens, D. Faraday waves on band pattern under zero gravity conditions. *Phys. Rev. Fluids* **4**, 064001 (2019).

28. Beysens, D., Chatain, D., Evesque, P. & Garrabos, Y. High-Frequency Driven Capillary Flows Speed Up the Gas-Liquid Phase Transition in Zero-Gravity Conditions. *Phys. Rev. Lett.* **95**, 034502 (2005).

29. Mialdun, A., Ryzhkov, I. I., Melnikov, D. E. & Shevtsova, V. Experimental Evidence of Thermal Vibrational Convection in a Nonuniformly Heated Fluid in a Reduced Gravity Environment. *Phys. Rev. Lett.* **101**, 084501 (2008).

30. Fernández, J., Tinao, I., Porter, J. & Laverón-Simavilla, A. Instabilities of vibroequilibria in rectangular containers. *Physics of Fluids* **29**, 024108 (2017).

31. Krieger, M. S. Interfacial fluid instabilities and Kapitsa pendula. *Eur Phys J E Soft Matter* **40**, 67 (2017).



\maketitle
\newpage
\section{Summary}

We aim to explain quantitatively the experimental results presented in the main text. We first model our system as a semi-circular solid pendulum shaken in an arbitrary direction. The velocity of the liquid interface will be identified with the velocity of the flat part of the pendulum. This model reproduces with good agreement all experimental data. Moreover, we show that the velocity found in our solid approach can be interpreted as an equilibrium condition for a free liquid interface. In the compressible case, the solid approach could not be carried as the liquid slab is now deformable but the dynamic equilibrium condition still holds. We will also discuss the velocity near the walls as the boundary layer plays a major role in our system.

The supplementary material is organized as follow. In \S 2, we briefly re-derive the equilibrium positions of a shaken pendulum. The \S 3 is dedicated to the mapping of the liquid on the equivalent semi-circular solid and to the introduction of notations. In particular, we give the correspondence between the quantities introduced in \S 2 and the quantity measured experimentally . We then derive in \S 4 the velocity field in the solid pendulum. Boundary layers are introduced in \S 5 as a perturbation of the velocity profile found for the solid. In \S 6, we compute the velocity of the interface starting from Navier-Stokes equation. When linearized, this expression is shown to be consistent with the velocity profile found in \S 4.      
\section{Pendulum shaken in arbitrary direction}
\begin{figure}[h!]
\center
\includegraphics[width=4cm]{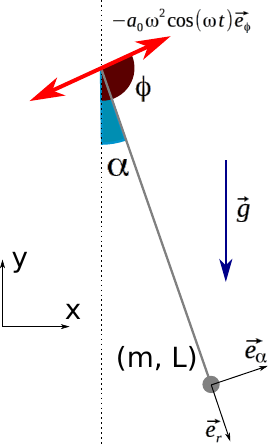}
\caption{Pendulum shaken with arbitrary acceleration}
\end{figure}
\paragraph{}We study the motion of a pendulum shaken with a speed $\vec{v} = -a_0 \omega \sin{(\omega t)} \vec{e}_\phi$ and an acceleration $\vec{a} = -a_0 \omega^2 \cos{(\omega t)}\vec{e}_\phi$ as presented in figure 1. The acceleration of the mass in the comoving frame is $L\ddot{\alpha} \et - L\dot{\alpha}^2 \er$ so that we have
\begin{equation}
m L\ddot{\alpha} = m \vec{g}.\et - m \vec{a}.\et 
\end{equation}
Defining $\omega_0^2 = g/L$ and $\xi = \frac{a_0 \omega^2}{g}$, this can be rewritten
\begin{equation}
\ddot{\alpha} + \omega_0^2 \sin{\alpha} = \xi \omega_0^2 \cos{(\omega t)} \sin{(\phi - \alpha)}
\label{eq:pfd}
\end{equation}
Assuming $\omega >> \omega_0$ we decompose $\alpha = \alpha_s + \alpha_f$ where $\alpha_f$ is a fast variable oscillating at $\omega$ and $\alpha_s$ is a slow variable. If $<.>$ is the mean over one fast period, one has $<\alpha_f> = <\alpha_s \cos{(\omega t)}> = 0$. We also assume that $\alpha_f$ is small. Keeping only leading order oscillating at $\omega$ in equation \ref{eq:pfd} leads to
\begin{equation}
\ddot{\alpha_f} = \xi \omega_0^2 \cos{(\omega t)} \sin{(\phi - \alpha_s)}
\end{equation}
that gives
\begin{equation}
\alpha_f = -\xi \frac{\omega_0^2}{\omega^2} \cos{(\omega t)} \sin{(\phi - \alpha_s)}
\label{eq:fast}
\end{equation}
We then take the mean $<.>$ of equation \ref{eq:pfd} and we get
\begin{equation}
\begin{split}
\ddot{\alpha_s} + \omega_0^2 \sin{\alpha_s} &= \xi \omega_0^2 <\cos{(\omega t)} [\sin{(\phi - \alpha_s)} - \cos{(\phi - \alpha_s)} \alpha_f ]> \\
&= - \xi \omega_0^2 \cos{(\phi - \alpha_s)} <\cos{(\omega t)}\alpha_f> \\
&= \xi^2 \frac{\omega_0^4}{2 \omega^2} \cos{(\phi - \alpha_s)} \sin{(\phi - \alpha_s)} \\
&= \xi^2 \frac{\omega_0^4}{4 \omega^2} \sin{(2\phi - 2\alpha_s)} 
\end{split}
\label{eq:pend}
\end{equation}
The equilibrium position then verifies

\begin{equation}
\xi^2 \frac{\omega_0^2}{4 \omega^2} \sin{(2\phi - 2\alpha_s)} = \sin{(\alpha_{eq})}
\label{eq:eqCondition}
\end{equation}

In the case of large excitation, we have $\xi >> 1$ so that $E_p$ is minimal for $\alpha_s \approx \phi + n\pi$ with $n$ an integer. We write $\phi + n\pi - \alpha_s = \epsilon <<1$ and we rewrite eq. \ref{eq:pend} as
\begin{equation}
- \ddot \epsilon + \omega_0^2 [ \sin{(\phi + n\pi)} - \cos{(\phi+n\pi)} \epsilon ] = \xi^2 \frac{\omega_0^4}{2\omega^2} \epsilon
\end{equation}
that can be reorganized as
\begin{equation}
\ddot \epsilon + \omega_0^2 \left( \cos{(\phi + n\pi)} + \frac{\xi^2 \omega_0^2}{2 \omega^2} \right) \epsilon = \omega_0^2 \sin{(\phi + n\pi)}
\end{equation}
There are two equilibrium positions. If we set $\phi = 0$ the equilibrium positions are $\epsilon = 0$ (hanging pendulum, $n=0$) and $\epsilon = \pi$ (inverted pendulum, $n=1$). In our experiments, we slowly vary $\phi$ and start with $\epsilon = 0$. For each angle, we can only observe the equilibrium that corresponds to $n=0$ as observing the equilibrium associated to $n=1$ would require to "jump" from one equilibrium to another. The equation of motion is then
\begin{equation}
\ddot \epsilon + \omega_0^2 \left( \cos{(\phi)} + \frac{\xi^2 \omega_0^2}{2 \omega^2} \right) \epsilon = \omega_0^2 \sin{(\phi)}
\end{equation}
that is the equation used in the main text.

\section{Mapping on the fluid interface experiment}
\subsection{Correspondance of variables}
\begin{figure}
\center
\includegraphics[width=9cm]{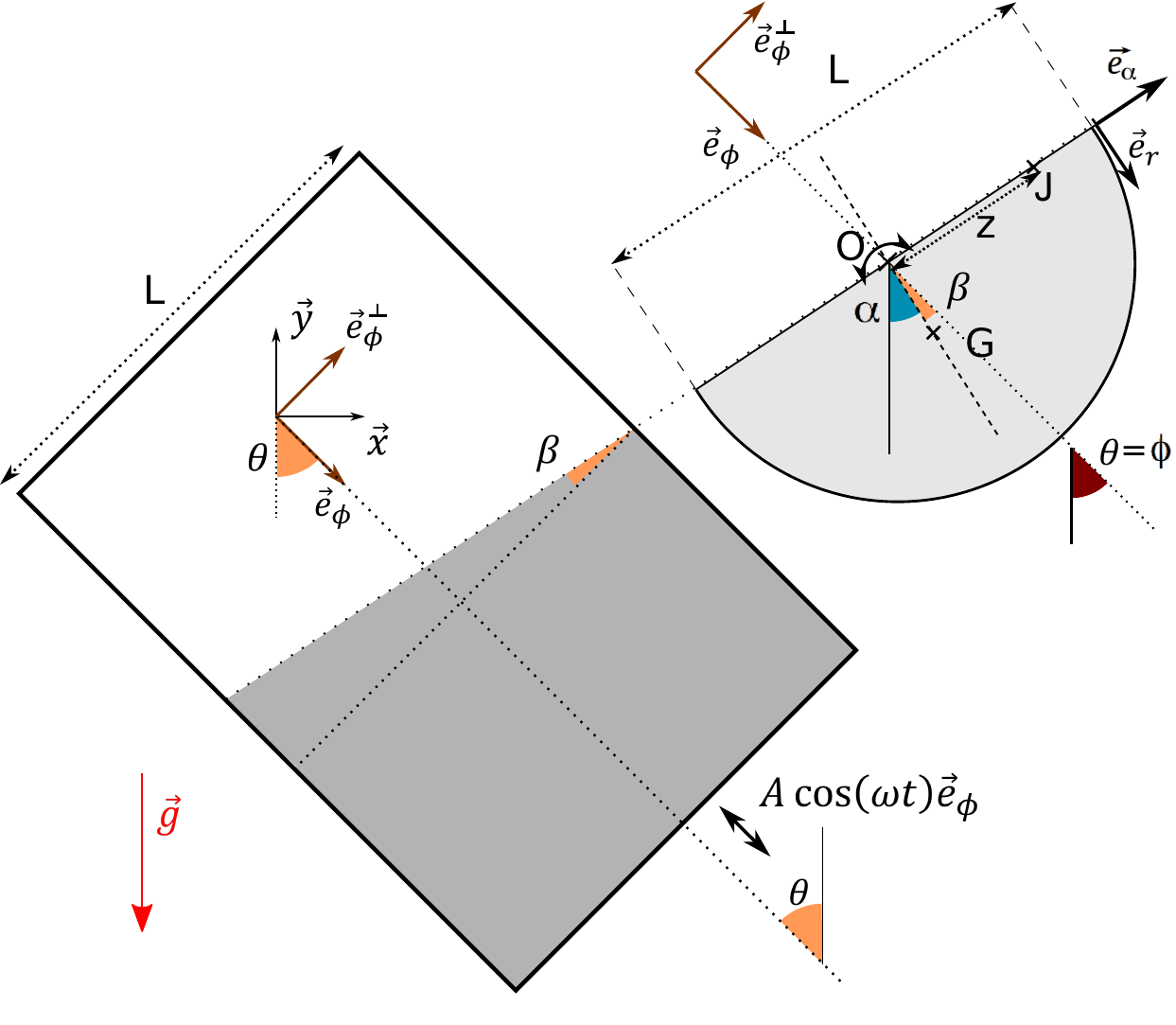}
\caption{Notations for the liquid and the equivalent disk with notations from the pendulum}
\end{figure}
\paragraph{}We want to link the angles of the pendulum $(\alpha_{eq}, \phi)$ with the angles defined for the fluid in the main text $(\beta, \theta)$. We first have $\theta = \phi$ (see figure 2). Concerning $\beta$, we see figure 2 that we have $\beta = \phi - \alpha_{eq}$.  This angle is zero when the interface is orthogonal to the shaking and is slightly positive otherwise.

\paragraph{}In order to make an analogy with a pendulum, we need to compute the equivalent $\omega_0$ of the effective pendulum. For an horizontal interface, we know that the largest mode on an interface of length L has a pulsation  $\sqrt{\frac{2 \pi g}{L}}$ . This has no reason to be valid for a non-horizontal interface. However, if we now consider our mass of fluid as a rotating half disk of mass m around it center, the inertia momentum is $I = \frac{\pi}{4} R^4 \rho  l_z = \frac{1}{2} m R^2$ and the center of mass is at a distance $\frac{4}{3 \pi} R$. The equation of motion for such disk is (with L= 2R)
\begin{equation}
\ddot{\alpha} = - m g \frac{4R}{3 \pi I} \sin{\alpha} = - \frac{16 g}{3 \pi L} \sin{\alpha} = -\omega_0^2 \sin{\alpha} 
\end{equation}
The main difference is that this model takes into account the motion of masses of liquids (sloshing) while the wave dispersion only takes into account surface deformation. Assuming rotation around point O ensures that the volume of fluid will be conserved.

Although the scaling of $\omega_0 \sim \sqrt{g/L}$ seems reasonable, the prefactor can be adjusted. In everything that follows we take $\omega_0 = \sqrt{\pi g/L}$ to give it a wave equivalent. This value ensures good agreement between experimental results and predictions. 

\subsection{Notations}
The velocity in the lab frame is noted $v(z, t) = V(z) \sin{(\omega t)}$, the velocity in the co-moving frame is noted $v^*(z, t) = V^*(z) \sin{(\omega t)}$ and the velocity in the boundary layer in the co-moving frame is noted $v_b^*(z, t) = V_b^*(z) \sin{(\omega t)}$

\section{Velocity profile of the interface}
We are now interested in the velocity of the interface. We take a point J such that in the co-moving frame $\vec{OJ} = z \et$. We have in the co-moving frame
\begin{equation}
\vec{v}^* (z, t) = - z \dot{\alpha} \er
\end{equation} 
We know that $\dot{\alpha} = \dot{\alpha_s} + \dot{\alpha_f}$. From equation \ref{eq:fast} we get
\begin{equation}
\dot{\alpha_f} = \xi \frac{\omega_0^2}{\omega} \sin{(\omega t)} \sin{(\phi - \alpha_s)}
\end{equation}
Moreover we assume that the equilibrium for slow variables is reached so that $\dot \alpha_s \approx 0$.
 Thus we get that in the co-moving frame
\begin{equation}
\vec{v}^* (z, t) = - \xi z \frac{\omega_0^2}{\omega} \sin{(\omega t)} \sin{(\phi - \alpha_{eq})} \er = V^*(z) \sin{(\omega t)} \er
\label{eq:vDisk}
\end{equation}
In the lab frame the velocity of the point J is (using $\er = \cos{(\phi - \alpha_{eq})} \vec{e}_\phi + \sin{(\phi - \alpha_{eq})} \vec{e}_\phi^\perp$)
\begin{equation}
\begin{split}
\vec{v}(z, t) &= -a_0 \omega \sin{(\omega t)} \vec{e}_{\phi} - \xi z \frac{\omega_0^2}{\omega} \sin{(\omega t)} \sin{(\phi - \alpha_{eq})} \er \\
&=- \left(a_0 \omega + \xi z \frac{\omega_0^2}{2 \omega} \sin{(2\phi - 2\alpha_{eq})}  \right) \sin{(\omega t)}  \vec{e}_{\phi} \\
& - \xi z \frac{\omega_0^2}{\omega} \sin{(\phi - \alpha_{eq})^2}  \sin{(\omega t)}\vec{e}_\phi^\perp \\
\end{split}
\label{eq:velocityLab}
\end{equation}
Experimentally we measure $|\beta| = |\phi-\alpha_{eq}| < \pi/6$ so that we can neglect the term along $\vec{e}_\phi^\perp$ in the previous expression. In particular we see that the point O has the velocity of the excitation.

\section{Boundary layer}

\begin{figure}[h!]
\centering
\includegraphics[width=10cm]{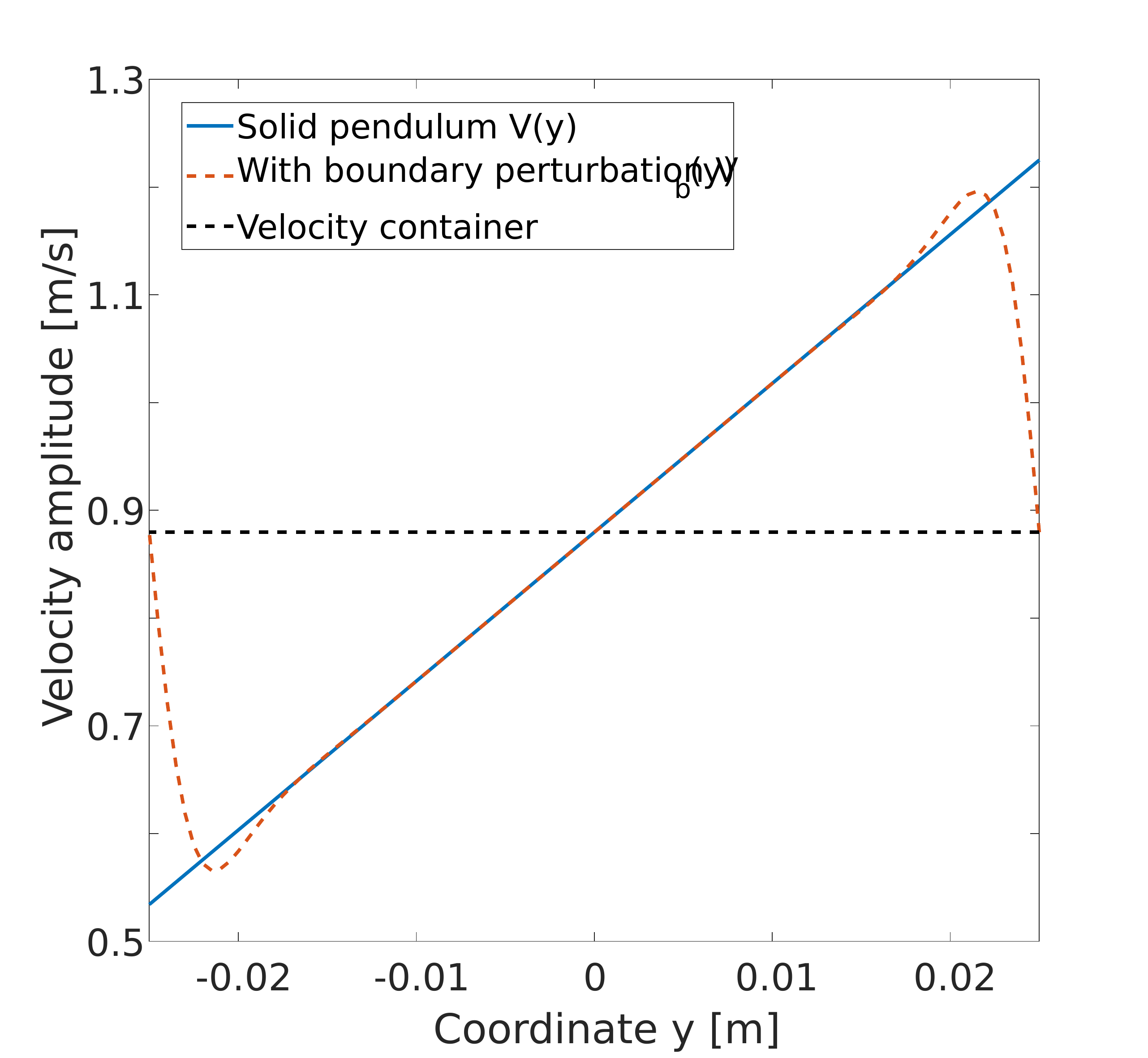}
\caption{Predicted velocity profile with typical values $a_0 = 1.4$ mm, $L = 5$ cm, $\omega = 2\pi . 100$ rad/s, $\xi = 55$ : blue line is the velocity profile from equation \ref{eq:velocityLab}, red dashed line is the velocity profile when the boundary layer is added. Black dashed line is the velocity of the container.}
\end{figure}

\paragraph{}We have seen in the previous section that the velocity is at first order aligned with the excitation, that means $\er \approx \vec{e}_\phi$. We will also use complex variables as our equations will be linear. Any velocity will take the form $\vec{v} (z, t) = V(z) e^{i \omega t} \vec{e}_\phi$. For instance the velocity at the interface found in equation \ref{eq:velocityLab} now writes
\begin{equation}
\begin{split}
\vec{v}(z, t) &= V(z) e^{i \omega t}  \vec{e}_{\phi} \\
V(z) &= - a_0 \omega - \xi z \frac{\omega_0^2}{2 \omega} \sin{(2\phi - 2\alpha_{eq})} \\
& = - a_0 \omega + V^*(z)
\end{split}
\end{equation}

Due to symmetry, all variables are assumed to depend on $z$ (and $t$) only. If we assume non-slip boundary condition at the walls, the velocity should verify 
\begin{equation}
V(z = \pm L/2) = -a_0 \omega 
\end{equation}
Clearly the velocity $V(z)$ does not verify this condition (see figure 3). We will perturb it near the wall to satisfy this condition.  The spatial extension of the perturbation should be of the order of $\delta = \sqrt{\frac{2 \nu}{\omega}} \approx 2$ mm. In this boundary layer, the flow $\vec v_b$ should follow Stokes equation with boundary conditions
\begin{equation}
\begin{split}
&\frac{\partial \vec{v}_b}{\partial t} = \nu \Delta \vec{v}_b - \frac{1}{\rho} \vec \nabla P \\
& \vec{v}_b(z=\pm L/2) = -a_0 \omega e^{i \omega t} \vec e_\phi   \\
& \vec{v}_b(|z \pm L/2| >> \delta) = V(z) e^{i \omega t} \vec e_\phi.
\label{eq:stokes}
\end{split}
\end{equation}
The last conditions indicates that the velocity at a distance $d >> \delta$ from the wall is simply the unperturbed velocity. Over a few $\delta$, the velocity $V(z)$ can be considered as a constant (see figure 3). The variation over $\delta$ is $\frac{dV}{dz} \delta = a_0 \omega \sin{(\beta)}  \frac{\delta}{2L}$ so that compared to the typical velocity $V \approx a_0 \omega$ one can neglect this variation as $\delta / L <<1$. This fact will now be used to find the velocity profile near the wall.

To solve equation \ref{eq:stokes} we use superposition principle. We first consider the flow $\vec{v}_\infty = V(\pm L/2) e^{i\omega t} \vec e_\phi$. As $\Delta \vec{v}_\infty$ = 0, the gradient pressure reads $\frac{1}{\rho} \vec \nabla P_\infty = -i \omega \vec{v}_\infty$.  We then define $ \vec{w}_b = \vec v_b - \vec v_\infty$ and $\hat P = P - P_\infty$. One can show that these new variables obey the system
\begin{equation}
\begin{split}
& \vec \nabla \hat P = 0 \\
&\frac{\partial \vec{w}_b}{\partial t} = \nu \Delta \vec{w}_b \\
& \vec{w}_b(\pm L/2) = -V^*(\pm L/2) e^{i \omega t} \vec e_\phi   \\
& \vec{w}_b(|z \pm L/2| >> \delta) \approx 0.
\label{eq:stokes2}
\end{split}
\end{equation}
where we used for the last condition that V(z) is approximately constant in the boundary layer. The solution to this equation is (writing $\vec{w}_b(z, t) = W_b(z) e^{i \omega t} \vec e_\phi$ )
\begin{equation}
W_b =  - V^*(\pm L/2) e^{-\frac{1+i}{\delta}(\frac{L}{2} - |z|)}
\end{equation} 
where as expected the size of the boundary layer is
\begin{equation}
\delta = \sqrt{\frac{2 \nu}{\omega}}
\end{equation}
At the end we get the corrected velocity profile
\begin{equation}
V_b(z) = V(z) - V^*(\pm L/2) e^{-(1+i) \phi(z)}
\label{eq:vLab}
\end{equation}
where $\phi(z) = \frac{1}{\delta}(\frac{L}{2} - |z|)$ is the distance from one wall divided by $\delta$. We see that few $\delta$ away from the wall, we recover as expected the velocity computed in the previous section. In contrary, at $z=\pm L/2$ the velocity is $-a_0 \omega $ that is exactly the velocity of the container.

\section{Equilibrium condition from Navier-Stokes equation}
\begin{figure}
\centering
\includegraphics[width=8cm]{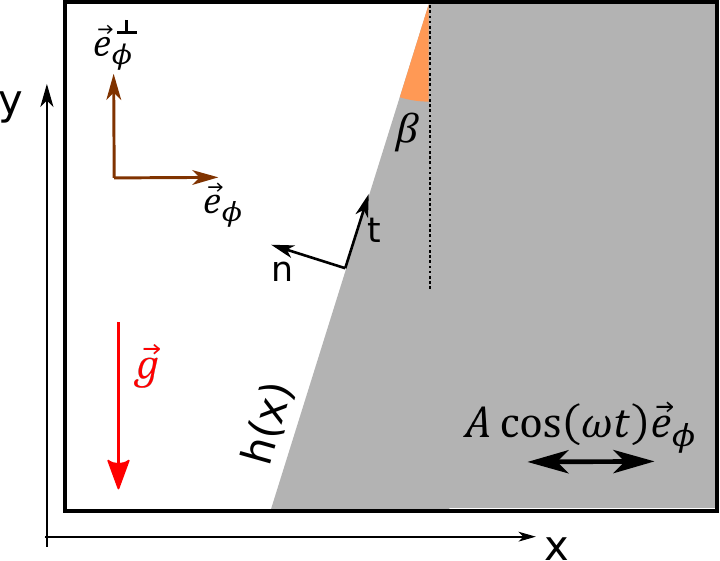}
\caption{Sketch of the liquid and notations}
\end{figure}

We now propose to write the equilibrium condition at the free interface of a vibrated liquid. We will assume that the forcing is horizontal so that $\vec e_\phi = \vec e_x$. We start from Navier-Stokes equation in the comoving frame of velocity $-a_0 \omega \sin{(\omega t)} \vec e_x$ that writes
\begin{equation}
\begin{split}
\vec \nabla . \vec v^* &= 0 \\
\frac{\partial \vec v^*}{\partial t}+\vec v^* . \vec \nabla \vec v^* &= -\frac{1}{\rho}\vec \nabla p + \nu \Delta \vec v^* + \vec g + a_0 \omega^2 \cos{(\omega t)} \vec e_x 
\end{split}
\label{eq:NS}
\end{equation}
At the free boundary of equation $y = h(x, t)$ in the comoving frame one has the kinematic condition and the dynamic condition
\begin{equation}
\begin{split} 
\frac{\partial h}{\partial t} + v^*_x \frac{\partial h}{\partial x} &= \frac{dh}{dt} = v_y^* \\
p(x, h(x, t)) &= P_{atm} \\
\label{eq:boundary}
\end{split}
\end{equation}
where $v_n$ is the velocity normal to the surface and $P_{atm}$ is the atmospheric pressure. This holds providing we neglect capillary effect and variation of pressure in the air layer. We will expand the quantities in the same manner as in reference [8]. We assume that $a << a \omega << a \omega^2$ and that $a \omega^2 >> g$. We define the different time scales as
\begin{equation}
\begin{split}
t_{-1} &= \omega t \\
t_0 &= t \\
t_1 &= t/\omega ...
\end{split}
\end{equation}
and we expand all quantities as
\begin{equation}
\begin{split}
p & = \omega p_{-1}(t_{-1}, t_0, ...) + p_0(t_{-1}, t_0, ...) + ... \\
h & = h_0(t_{-1}, t_0, ...) + \frac{1}{\omega} h_1(t_{-1}, t_0, ...) \\
\frac{\partial \vec v^*}{\partial t} &= \omega \frac{\partial \vec v^*}{\partial t_{-1}} +  \frac{\partial \vec v^*}{\partial t_0} + ...
\end{split}
\end{equation}
Inserting these variables in eq. \ref{eq:NS} gives at leading order in $\omega$
\begin{equation}
\frac{\partial \vec v^*}{\partial t_{-1}} = -\frac{1}{\rho}\vec \nabla p_{-1} +  a_0 \omega \cos{(t_{-1})} \vec e_x
\end{equation}
suggesting that all quantities oscillates with respect to $t_{-1}$. Writing $\vec v^* = \vec V^*(t_0) \sin{(t_{-1})} + \vec u^* (t_0)$ and $p_{-1} = P_{-1}(t_0) \cos{(t_{-1})}$ we get
\begin{equation}
\vec \nabla P_{-1} = - \rho \left(\vec V^* - a_0 \omega \vec e_x \right) = -\rho \vec V
\label{eq:P-1}
\end{equation}
Hence at leading order the pressure gradient is linked to the velocity in the laboratory frame $\vec V$. The term $\vec u^*$ corresponds to slow flow compared to $\omega$. As we are looking for equilibrium solutions for large times, we will assume that $\vec u^* = 0$ so that we have $\vec v^* = \vec V^*(t_0) \sin{(t_{-1})}$.

We now take the next leading order of equation \ref{eq:NS} and get
\begin{equation}
\frac{\partial \vec v^*}{\partial t_0} + \vec v^* . \vec \nabla \vec v^* = -\frac{1}{\rho}\vec \nabla p_0 + \nu \Delta \vec v^* + \vec g
\label{eq:P0}
\end{equation}
By taking the rotational of eq. \ref{eq:P-1} we show that $\vec \nabla \times \vec v^* = 0$. The non linear term can then be written as $ \vec v^* . \vec \nabla \vec v^* = \frac{1}{2}\vec \nabla (v^{*2})$. Taking the mean over one fast period of equation \ref{eq:P0} gives
\begin{equation}
\frac{1}{4}\vec \nabla (V^{*2}) = -\frac{1}{\rho}\vec \nabla \bar{p}_0 + \vec g
\end{equation}
where $\bar{p}_0$ is the mean pressure over one fast period. From this we deduce that the static pressure field in the fluid is
\begin{equation}
\bar{p}_0(x, y) = C_0 - \frac{1}{4}\rho V^{*2}(x, y) - \rho g y
\end{equation}

We will now look at the boundary conditions \ref{eq:boundary}. We will first compute $h(x, t)$. At first order, the kinematic condition is
\begin{equation}
\frac{\partial h_0}{\partial t_{-1}} = 0
\end{equation}
meaning that $h_0$ does not depend on $t_{-1}$. The next leading order gives ($\frac{\partial h_0}{\partial t_0} = 0$ at equilibrium)
\begin{equation}
\frac{\partial h_1}{\partial t_{-1}} + V_{x}^* \frac{\partial h_0}{\partial x} \sin{(t_{-1})} = V_y^* \sin{(t_{-1})}
\end{equation}
From this we get
\begin{equation}
h(x, t) = h_0(x) + \frac{1}{\omega} \left(V_{x}^* \frac{\partial h_0}{\partial x} - V_{y}^* \right) \cos{(\omega t)}
\end{equation}

We can now develop the dynamic condition as
\begin{equation}
\omega p_{-1} (x, h_0) + \omega \frac{\partial p_{-1}}{\partial y} (x, h_0) \frac{1}{\omega} \left(V_{x}^* \frac{\partial h_0}{\partial x}  - V_{y}^* \right) \cos{(\omega t)} + p_0(x, h_0) = P_{atm} - C
\end{equation}
At first order we get $\omega p_{-1}(x, h_0) = C_{-1}$. This conditions forces the gradient of pressure in the direction tangent to the interface to be zero. This gives using equation \ref{eq:P-1}
\begin{equation}
\vec{V} . \vec t = 0
\end{equation}  
At the interface, the velocity amplitude can then simply be written
\begin{equation}
\vec V = V(x, y) \vec n
\end{equation}

We have $\frac{\partial p_{-1}}{\partial y} = \frac{\partial P_{-1}}{\partial y} \cos{(\omega t)} = - \rho V_{y} \cos{(\omega t)} $ using equation \ref{eq:P-1}. Taking the mean over one fast period gives (using the computed expression for $\bar p_0$)
\begin{equation}
- \frac{1}{2} V_{y} \left(V_{x}^* \frac{\partial h_0}{\partial x}  - V_{y}^* \right) - \frac{1}{4}V^{*2}(x, h_0) - g h_0(x) = P_{atm}- C
\end{equation}

In order to go further, we will assume a linear profile for $h_0(x) = \frac{x-x_0}{\tan{\beta}}$ with $\beta > 0$ the equilibrium angle. Under this assumption the vector normal to the surface is $\vec n = - \cos{\beta}\vec e_x + \sin{\beta} \vec e_y$ and $\frac{\partial h_0}{\partial x}  = 1/\tan{\beta}$. We finally need the components of the velocity field that are 
\begin{equation}
\begin{split}
\vec V &= V (- \cos{\beta}\vec e_x + \sin{\beta} \vec e_y) \\
V_{y} &= \sin{\beta} V \\
\vec V^* &= a_0 \omega  \vec e_x + \vec V \\
V_{x}^* &= a_0 \omega - V \cos{\beta} \\
V_{y}^* &= \sin{\beta} V\\
V^{*2} &= V_x^{*2} + V_y^{*2}
\end{split}
\end{equation}
This finally gives
\begin{equation}
\frac{1}{4}  V^{*2} - \frac{1}{2} a_0 \omega V_x^* - gh_0(x) = P_{atm}- C 
\end{equation}

In the limit of large forcing, we expect $\beta << 1$ so that  $V_x^{*2} \approx V^{*2}$. In order to conserve volume we impose $\int V^*(x)dx = 0$. We deduce that there is at least one point $x_0$ such that $V^*(x_0) = 0$ since $V^*$ is expected to be continuous. We evaluate the previous expression at this point to get the constant $P_{atm}- C = -gh_0(x_0) = -g y_0$. From this we finally get
\begin{equation}
V^*(V^* - 2 a_0 \omega) = 4g(y-y_0)
\label{eq:dynEq}
\end{equation}

Around the point $y_0$ we have $V^* << 2 a_0 \omega$ so that 
\begin{equation}
\begin{split}
v^* (y, t) & \approx -\frac{2g}{a_0 \omega} (y-y_0) \sin{(\omega t)} \\
& = -\frac{2}{\xi} \omega (y - y_0) \sin{(\omega t)}
\end{split}
\end{equation}

We can compare this to the velocity of the rigid pendulum that was (see equation \ref{eq:vDisk} with $z= \frac{y-y_0}{\sin{(\alpha_{eq})}}$)
\begin{equation}
v^*(y, t) = - \xi \frac{y-y_0}{\sin{(\alpha_{eq})}} \frac{\omega_0^2}{\omega}  \sin{(\phi - \alpha_{eq})} \sin{(\omega t)}
\end{equation}
Both velocities are equal if $\sin{(\alpha_{eq})} =  \xi^2 \frac{\omega_0^2}{2 \omega^2} \sin{(\phi - \alpha_{eq})}$. At first order in $\phi-\alpha_{eq}$ this is exactly the equilibrium condition \ref{eq:eqCondition}. In the limit of large forcing, the velocity found in the solid case can be interpreted as the velocity of the interface satisfying the equilibrium condition.

Note that since $V^2 = (a_0 \omega - V^*)^2 = a_0^2 \omega^2 + V^* (V^* - 2 a_0 \omega)$ the condition \ref{eq:dynEq} can also be written 
\begin{equation}
V^2 = a_0^2 \omega^2 + 4g(y-y_0)
\end{equation}
The point $y_0$ being determined by the volume conservation condition. This gives
\begin{equation}
V(y) = \sqrt{a_0^2 \omega^2 + 4g(y-y_0)}
\label{eq:vBernFinal}
\end{equation}
as long as $a_0^2 \omega^2 > 4g(y-y_0)$. Thus the maximum height that can be stabilized is related to the forcing velocity.

\end{document}